\documentclass[11pt,letterpaper]{llncs}

\usepackage{wrapfig}
\usepackage{authblk}
\usepackage{listings}
\usepackage{color}
\usepackage{amsmath}  
\usepackage{graphicx} 
\usepackage{amssymb} 
\usepackage{algpseudocode}
\usepackage{algorithm}
\usepackage{comment}
\usepackage{hyperref}
\usepackage{orcidlink}
\usepackage{amsfonts}
\usepackage{float}
\floatstyle{plaintop}
\restylefloat{table}
\usepackage{fancyhdr}

\usepackage{geometry}
\newcommand{\keywords}[1]{\par\addvspace\baselineskip
\noindent\keywordname\enspace\ignorespaces#1}

\usepackage{listings}
\usepackage{color}
\usepackage{amsmath}  
\usepackage{graphicx} 
\usepackage{amssymb} 
\usepackage{algpseudocode}
\usepackage{algorithm}
\usepackage{comment}
\usepackage{hyperref}
\usepackage{orcidlink}

\usepackage{float}
\floatstyle{plaintop}
\restylefloat{table}

\begin{document}

\title{\LARGE{Privacy-aware White and Black List Searching for Fraud Analysis}}

\author{\large{William J Buchanan\inst{1}, Jamie Gilchrist\inst{2}, Zakwan Jaroucheh\inst{2}, Dmitri Timosenko\inst{2}, Nanik Ramchandani\inst{2}, Hisham Ali\inst{1}}}
\institute{
\large{{Blockpass ID Lab, Edinburgh Napier University, Edinburgh.}
\and {LastingAsset, 6 Bainfield Drive, C/o Bright Red Triangle, Edinburgh}} }

\maketitle

\begin{abstract} 
In many areas of cybersecurity, we require access to Personally Identifiable Information (PII), such as names, postal addresses and email addresses. Unfortunately, this can lead to data breaches, especially in relation to data compliance regulations such as GDPR. An Internet Protocol (IP) address is an identifier that is assigned to a networked device to enable it to communicate over networks that use IP. Thus, in applications which are privacy-aware, we may aim to hide the IP address while aiming to determine if the address comes from a blacklist. One solution to this is to use homomorphic encryption to match an encrypted version of an IP address to a blacklisted network list. This matching allows us to encrypt the IP address and match it to an encrypted version of a blacklist. In this paper, we use the OpenFHE library \cite{OpenFHE} to encrypt network addresses with the BFV homomorphic encryption scheme. In order to assess the performance overhead of BFV, we implement a matching method using the OpenFHE library and compare it against partial homomorphic schemes, including Paillier, Damgard-Jurik, Okamoto-Uchiyama, Naccache-Stern and Benaloh. The main findings are that the BFV method compares favourably against the partial homomorphic methods in most cases. 
\end{abstract}


\section{Introduction}
Data regulations such as GDPR demand greater control of Personally Identifiable Information (PII). In many areas of cybersecurity, we provide linkages between entities and their associated IP address and where revealing an IP address can often identify a person or organisation that is involved in an investigation. With this, we could define a blacklist of networks that we need to identify if specific source IP address is included. One of the best ways to preserve privacy is with the use of homomorphic encryption, where we can encrypt both the target IP address and the blacklist and match them without revealing any additional information. 

Homomorphic encryption allows us to take the plaintexts $m_1$ and $m_2$ encrypt them using a secret key $k$, and perform operations such that:

\[
    \mathsf{Enc}_k(m_1 \circ m_2) = \mathsf{Enc}_k(m_1) \circ \mathsf{Enc}_k(m_2)  
\]
where $\circ$ could potentially be any operator, such as add, multiply, logical and, or logical or. With symmetric key encryption, we use the same key to decrypt as we do to encrypt. Overall, in analysing the IP matching problem, we need to either conduct bitwise homomorphic encryption or use a homomorphic subtraction method. 

In the past, partial homomorphic methods (PHE) have been used within privacy-aware methods for network analysis. This includes Tusa \emph{et al.}, who used the Paillier method to implement privacy-aware routing  \cite{tusa2023homomorphic}. These methods often have fairly good performance levels, but they do not implement a full range of mathematical operations and thus often fail to scale on a large-scale basis, especially where the full range of operations is required. This paper thus provides a new method for the usage of fully homomorphic encryption to match IP addresses to a blacklist of network addresses without revealing the IP address or the blacklist.

\section{Fully Homomomorphic Encryption}

Homomorphic encryption is a method of encryption which supports operations over encrypted data. In 1978, Rivest, Adleman, and Dertouzos \cite{rivest1978data} were the first to explore the possibilities of using the natural homomorphic properties of the RSA public key encryption scheme. The RSA scheme only supports the evaluation of arithmetic multiplication over ciphertexts. RSA is an example of Partial Homomorphic Encryption (PHE), which is a scheme that supports the evaluation of only a single type of operation on ciphertexts. Fully Homomorphic Encryption (FHE) can support every operation.  Since Gentry defined the first FHE method \cite{homenc} in 2009, there have been four main generations of homomorphic encryption:

\begin{itemize}
    \item 1st generation: Gentry’s method uses integers and lattices. \cite{van2010fully} including the DGHV method.
    \item 2nd generation. Brakerski, Gentry and Vaikuntanathan’s (BGV) and Brakerski/ Fan-Vercauteren (BFV) use a Ring Learning With Errors approach \cite{brakerski2014efficient}.  The methods are similar to each other and have only a small difference between them.
    \item 3rd generation: These include DM (also known as FHEW) and CGGI (also known as TFHE) and support the integration of  Boolean circuits for small integers. 
    \item 4th generation: CKKS (Cheon, Kim, Kim, Song) and which uses floating-point numbers \cite{cheon2017homomorphic}.
\end{itemize}

\subsection{Public key or symmetric key}
\begin{figure*}
\begin{center}
  \includegraphics[width=0.9\linewidth]{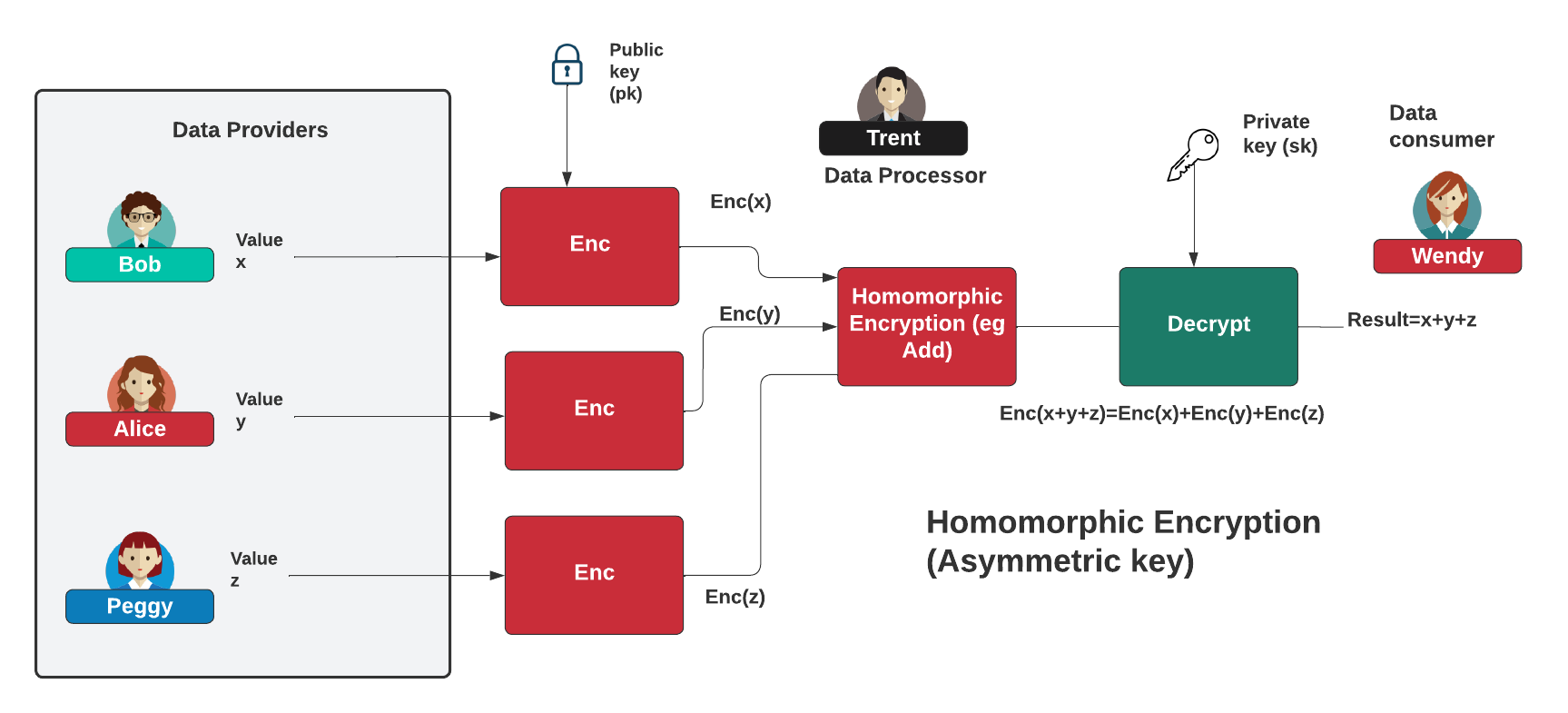}
  \caption{Asymmetric encryption (public key)}
  \label{fig:asym}
  \end{center}
\end{figure*}

Homomorphic encryption can be implemented either with a symmetric key or an asymmetric (public) key. With symmetric key encryption, we use the same key to encrypt as we do to decrypt, whereas, with an asymmetric method, we use a public key to encrypt and a private key to decrypt.  In Figure \ref{fig:asym}, we use asymmetric encryption with a public key ($pk$) and a private key ($sk$). With this, Bob, Alice and Peggy will encrypt their data using the public key to produce ciphertext, and then we can operate on the ciphertext using arithmetic operations. The result can then be revealed by decrypting with the associated private key. We can also use symmetric key encryption, where the data is encrypted with a secret key, and which is then used to decrypt the data. In this case, the data processor (Trent) should not have access to the secret key, as they could decrypt the data from the providers.

\subsection{Homomorphic libraries}
There are several homomorphic encryption libraries that support FHE, including ones that support CUDA and GPU acceleration, but many have not been kept up-to-date with modern methods or have only integrated one method. Overall, the native language libraries tend to be the most useful, as they allow the compilation to machine code. The main languages for this are C++, Golang and Rust, although some Python libraries exist through wrappers to C++ code. This includes HEAAN-Python and its associated HEAAN library.  

One of the first libraries which supported a range of methods are Microsoft SEAL \cite{asecuritysite_85691}, SEAL-C\# and SEAL-Python. While it supports a wide range of methods, including BGV/BFV and CKKS, it has lacked any real serious development for the past few years. Wood et al. \cite{wood2020homomorphic} define a full range of libraries. One of the most extensive libraries is PALISADE, and which has now developed into OpenFHE.  Within OpenFHE, the main implementations are:

\begin{itemize}
    \item Brakerski/Fan-Vercauteren (BFV) scheme \cite{asecuritysite_11288} for integer arithmetic.
    \item Brakerski-Gentry-Vaikuntanathan {BGV}) scheme for integer arithmetic
    \item Cheon-Kim-Kim-Song (CKKS) scheme for real-number arithmetic (includes approximate bootstrapping)
    \item Ducas-Micciancio (DM) and Chillotti-Gama-Georgieva-Izabachene (CGGI) schemes for Boolean circuit evaluation.
\end{itemize}

\subsection{Bootstrapping}
\begin{figure*}
\begin{center}
  \includegraphics[width=0.9\linewidth]{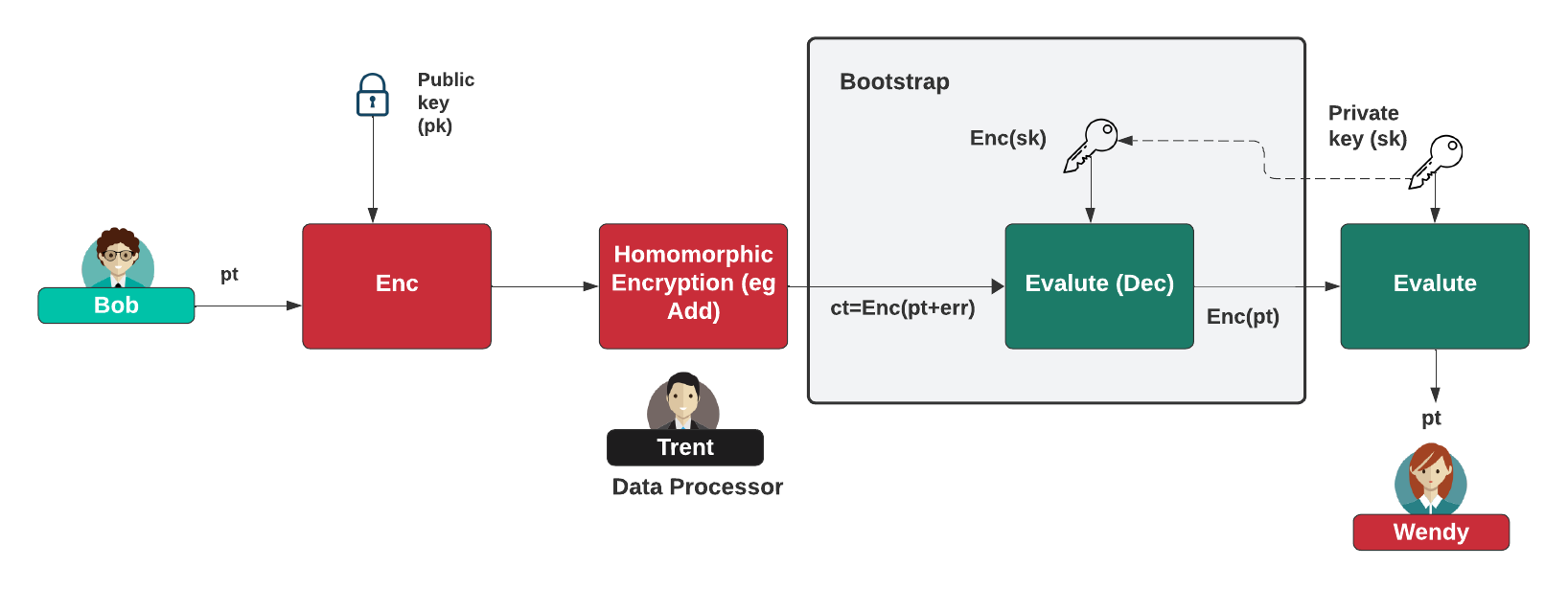}
  \caption{Bootstrap}
  \label{fig:bootstrap}
  \end{center}
\end{figure*}

A key topic within fully homomorphic encryption is the usage of bootstrapping. Within a learning with-errors approach, we add noise to our computations. For a normal decryption process, we use the public key to encrypt data and then the associated private key to decrypt it. Within the bootstrap version of homomorphic encryption, we use an encrypted version of the private key that operates on the ciphertext. In this way, we remove the noise which can build up in the computation. Figure \ref{fig:bootstrap} outlines that we perform an evaluation on the decryption using an encrypted version of the private key. This will remove noise in the ciphertext, after which we can then use the actual private key to perform the decryption.

The main bootstrapping methods are CKKS \cite{cheon2017homomorphic}, DM \cite{ducas2015fhew}/CGGI, and BGV/BFV. Overall, CKKS is generally the fastest bootstrapping method, while DM/CGGI is efficient with the evaluation of arbitrary functions. These functions approximate math functions as polynomials (such as with  Chebyshev approximation). BGV/BFV provides reasonable performance and is generally faster than DM/CGGI but slower than CKKS.

\subsection{BGV and BFV}
With BGV and BFV \cite{asecuritysite_11288}, we use a Ring Learning With Errors (LWE) method \cite{brakerski2014efficient}.  With BGV, we define a modulus  ($q$), which constrains the range of the polynomial coefficients. Overall, the methods use a moduli, which can be defined within different levels. We then initially define a finite group of $\mathbb{Z}_q$ and then make this a ring by dividing our operations with $(x^n+1)$ and where $n-1$ is the largest power of the coefficients. The message can then be represented in binary as:

\begin{equation}
m=a_{n-1}a_{n-2}...a_0
\end{equation}

This can be converted into a polynomial with:

\begin{equation}
\mathbf{m}=a_{n-1} x^{n-1} + a_{n-2} x^{n-2}+...+a_1 x + a_0 \pmod q
\end{equation}

The coefficients of this polynomial will then be a vector. Note that for efficiency, we can also encode the message with ternary (such as with -1, 0 and 1). We then define the plaintext modulus with:

\begin{equation}
t = p^r
\end{equation}

and where $p$ is a prime number and $r$ is a positive number. We can then define a ciphertext modulus of $q$, and which should be much larger than $t$. To encrypt with the private key of $\mathbf{s}$, we implement:

\begin{equation}
(c_0, c_1) =\left( \frac{q}{t}.\mathbf{m}  + \mathbf{a}.\mathbf{s} + e,\mathbf{-a} \right) \mod q
\end{equation}

To decrypt:

\begin{equation}
m = \bigl \lfloor \frac{t}{q}(c_0+c_1).\mathbf{s} \bigr \rceil
\end{equation}

\subsubsection{Noise and computation}
But each time we add or multiply, the error also increases. Thus, bootstrapping is required to reduce the noise. Overall, addition and plaintext/ciphertext multiplication is not a time-consuming task, but ciphertext/ciphertext multiplication is more computationally intensive. The most computational task is typically the bootstrapping process, and the ciphertext/ciphertext multiplication process adds the most noise to the process.

\subsubsection{Parameters}
We thus have a parameter of the ciphertext modulus ($q$) and the plaintext modulus ($t$). Both of these are typically to the power of 2. An example of $q$ is $2^{240}$ and for $t$ is 65,537. As the value of $2^q$ is likely to be a large number, we typically define it as a log\_q value.  Thus, a ciphertext modulus of $2^{240}$ will be 240 as defined as a log\_q value.

\section{Partial homomorphic encryption}
With partial homomorphic encryption (PHE), we can implement some form of arithmetic operation in a homomorphic way.  These methods include RSA, ElGamal, Paillier \cite{paillier1999public}, Exponential ElGamal, Elliptic Curve ElGamal \cite{elgamal1985public}, Paillier \cite{paillier1999public}, Damgard-Jurik \cite{damgaard2010generalization}, Okamoto–Uchiyama \cite{okamoto1998new}, Benaloh \cite{cohen1985robust}, Naccache–Stern \cite{naccache1997new}, and Goldwasser–Micali \cite{goldwasser2019probabilistic}. Overall, we can use RSA and ElGamal for multiplicative homomorphic encryption;  Paillier, Exponential ElGamal, Elliptic Curve ElGamal, Damgard-Jurik, Okamoto–Uchiyama, Benaloh and Naccache–Stern for additive homomorphic encryption; and Goldwasser–Micali for XOR homomorphic encryption.  

\subsection{Paillier}
The Paillier cryptosystem \cite{paillier1999public} is a partial homomorphic encryption (PHE) method that can perform addition, subtraction, and scalar multiplication. Thus we get:

\begin{align}
Enc_k(A + B) &= Enc_k(A) + Enc_k(B) \\
Enc_k(A - B) &= Enc_k(A) - Enc_k(B)  \\
Enc_k(A.B) &= A. Enc_k(B)  
\end{align}

If we take two values: $m_1$ and $m_2$, we get two encrypted values of $Enc(m_1)$ and $Enc(m_2)$. We can then multiply the two cipher values to get $Enc(m_1+m_2)$. We can then decrypt to get $m_1+m_2$. Along with this, we can also subtract to $Enc(m_1-m_2)$. This is achieved by taking the inverse modulus of $Enc(m_2)$ and multiplying it with $Enc(m_1)$. Finally, we can perform a scalar multiply to get $Enc(m_1 \cdot m_2)$ and which is generated from $Enc(m1)^{m2}$. 

First we select two large prime numbers ($p$ and $q$) and compute:
\begin{align}
N&=pq\\
PHI&=(p-1)(q-1)\\
\lambda &=\operatorname {lcm} {(p-1,q-1)} 
\end{align}

and where lcm is the least common multiple. We then select a random integer $g$ for:

\begin{align}
\displaystyle g \in \mathbb {Z} _{nN^{2}}^{*} 
\end{align}
We must make sure that $n$ divides the order of $g$ by checking the following: 
\begin{align}
\mu =(L(g^{\lambda }{\pmod {n}}^{2}))^{-1}{\pmod {N}}
\end{align}
and where $L$ is defined as $L(x)=\frac {x-1}{N}$. The public key is $(N,g)$, and the private key is $\lambda,\mu )$.

To encrypt a message ($M$), we select a random $r$ value and compute the ciphertext of:
\begin{align}
c=g^{m}\cdot r^{N} \pmod {N^{2}}
\end{align}

and then to decrypt:
\begin{equation}
m=L(c^{\lambda }{\pmod N}^{2})\cdot \mu {\pmod N}
\end{equation}

For adding and scalar multiplying, we can take two ciphers ($C_1$ and $C_2$), and get:
\begin{align}
C_1 = g^{m_1}\cdot r_1^{N} \pmod {N^{2}}\\
C_2 = g^{m_2}\cdot r_2^{N} \pmod {N^{2}}
\end{align}

If we now multiply them, we get:
\begin{align}
C_1 \cdot C_2 &= g^{m_1}\cdot r_1^{N} \cdot g^{m_2}\cdot r_2^{N} \pmod {N^{2}}\\
C_1 \cdot C_2 &= g^{m_1+m_2}\cdot r_1^{N} \cdot r_2^{N} \pmod {N^{2}}
\end{align}

\subsection{Benaloh}
In 1986, Josh (Cohen) Benaloh published on \emph{A Robust and Verifiable Cryptographically Secure Election Scheme} \cite{cohen1985robust,asecuritysite_15987}. Within it, Josh outlined a public key encryption method and where Bob could generate a public and a private key. Alice could then use the public key to encrypt data for Bob, and then Bob could use the associated private key to decrypt it. It has the advantage of supporting additive homomorphic encryption. For the Benaloh method, to generate a key pair, Bob generates $p$ and $q$, and which are two large prime numbers, which are two large distinct prime numbers. Next, he computes:

\begin{align}
n = pq\\
\phi(n) = (p – 1)(q – 1)
\end{align}

Bob then selects a block size ($r$) so that:

\begin{itemize}
    \item $r$ divides $p - 1$
    \item $\textrm{gcd}(r, (p - 1) / r) = 1$
    \item $\textrm{gcd}(r, q - 1) = 1$
\end{itemize}

Next Bob selects $y$ so that:

\begin{align}
x = y^{\phi(n) / r} \pmod n \neq 1 
\end{align}

Bob's private key is $(p, q)$, and his public key is  $(y, r, n)$. To encrypt data for Bob, Alice selects a message $(m$) and uses Bob's public key of $(y, r, n) $. First, she selects a random value of $u$ and which is between 0 and $n$.  Alice then encrypts with:

\begin{align}
c = y^m u^r \pmod n
\end{align}

She sends this ciphertext to Bob. He will then decrypt with:

\begin{align}
a = c^{\phi(n ) / r} \pmod n
\end{align}

Bob lets $md = 0$. If $x^{md} \pmod n \neq a$ the Bob increments $md$ by 1. He keeps doing this until:

\begin{align}
x^{md} \pmod n = a
\end{align}

The value of $md$ is then the original plaintext. One of the advantages of the Benaloh method is that we can perform additive homomorphic encryption. If we have two messages, we multiply the ciphers together for each message:

\begin{align}
c _1= y^{m_1} u^r \pmod n\\
c _2= y^{m_2} u^r \pmod n\\
c=c_1.c_2
\end{align}

This will give:
\begin{align}
c = y^{m_1+m_2} u^r \pmod n
\end{align}

\subsection{Okamoto-Uchiyama}
With the Okamoto-Uchiyama method \cite{okamoto1998new,asecuritysite_10221},  we can perform additive and scalar multiply homomorphic encryption. A public/private key pair is generated as follows:

\begin{itemize}
    \item Generate large primes p and q and set $n = p^2 q$.
    \item $ g\in ({\mathbb {Z}}/n{\mathbb {Z}})^{*}$ such that $g^{p-1}\neq 1\mod p^{2}$.
    \item Let $h = g^n \pmod n$.
\end{itemize}

 The public key is $(n, g, h)$, and then the private key is $(p, q)$. To encrypt a message m, where m is taken to be an element in $2^{k-1}$. We then select $ r\in \mathbb {Z} /n\mathbb {Z}$ at random. The cipher is then:
\begin{align}
C=g^{m}h^{r} \pmod n
\end{align}

Next, we define the function of: $L(x)=\frac{x-1}{p}$.

We then decrypt with:
\begin{align}
m={\frac {L\left(C^{{p-1}}\mod p^{2}\right)}{L\left(g^{{p-1}}\mod p^{2}\right)}} \pmod p
\end{align}

\subsection{Naccache–Stern}
With the Naccache–Stern method \cite{naccache1997new,asecuritysite_99911}, we select a large prime number ($p$). We then select a value ($n$) and for $i$ from 0 to $n$, we select the the first $n$ prime numbers ($p_0 ... p_{n-1}$ of which $p_0$ is 2. We must make sure that:

\begin{align}
\displaystyle \prod _{i=0}^{n}p_{i}&lt;p 
\end{align}

For our secret key ($s$) we make sure that:
\begin{align}
gcd(s,p-1)=1
\end{align}

To compute the public key ($v_i$), we calculate:
$ \displaystyle v_{i}={\sqrt[{s}]{p_{i}}}\pmod p $
To encrypt, we take a message of$m$ and then determine the message bits of$m_i$. We can then cipher with the public key:

\begin{align}
\displaystyle c=\prod _{i=0}^{n}v_{i}^{m_{i}}\pmod p
\end{align}

and then to decrypt:
\begin{align}
\displaystyle m=\sum _{i=0}^{n}{\frac {2^{i}}{p_{i}-1}}\times \left(\gcd(p_{i},c^{s}\pmod p)-1\right)
\end{align}

\subsection{Goldwasser–Micali}
With public key encryption, Alice could have two possible messages (a '0' or a '1') that she sends to Bob. If Eve knows the possible messages (a '0' or a '1'), she will then cipher each one with Bob's public key and then match the results against the cipher message that Alice sends. Eve can thus determine what Alice has sent to Bob. In order to overcome this, the Goldwasser–Micali (GM) method \cite{goldreich1987mental} is used as a public key algorithm that uses a probabilistic public-key encryption scheme. In this case, we will implement an XOR homomorphic encryption operation. For an input of 17 (10001) and 16 (10000), we will get a result of 00001 (1). 

In a probabilistic encryption method, Alice selects the plaintext ($m$) and a random string ($r$). Next, she uses Bob's public key to encrypt the message pair of $(m,r)$. If the value is random, then Eve will not be able to use the range of messages and random values used. If Bob wants to create his public and private keys. He first selects two random prime numbers for his private key and then calculates $N$:

\begin{align}
N=pq
\end{align}

The values of $p$ and $q$ will be his private key, and $N$ will form part of his public key. For the second part of his public key, he determines:

\begin{align}
a=\textrm{pseudosquare}(p, q)
\end{align}

For this, we determine if we can find, for a given value of $a$, which has no solutions: 
\begin{align}
u^2 \equiv a \pmod p\\
u^2 \equiv a \pmod q
\end{align}

This means that there are no quadratic residues. Bob's public key is $(N,a)$ and the private key is $(p,q)$. The key encryption method becomes:

\begin{itemize}
    \item Bob selects $p$ and $q$.
    \item Bob selects $a$ with $\left(\frac{a}{p}\right) $ = $\left(\frac{a}{q}\right) = -1$. This is a Jacobi symbol calculation.
    \item Bob publishes $N$ and $a$.
\end{itemize}

To encrypt for Bob:

\begin{itemize}
    \item Select a bit to encrypt $m \in {0,1}$.
    \item Alice uses Bob's values of $N,a)$ to compute:
    \item $c = r^2 \pmod N$ if $m=0$
    \item $c = ar^2 \pmod N$ if $m=1$
\end{itemize}

Alice chooses $r$ at random, and thus, Eve will not be able to spot the message, as the random values will consist of all possible squares modulo $N$ when $m=0$. Alice sends $c$ to Bob. To decrypt, Bob then computes $\left( \frac{c}{p}\right)$ and gets:
\begin{align}
m=0: if \left( \frac{c}{p}\right) = 1\\
m=1: if \left( \frac{c}{p}\right)  = -1
\end{align}
\subsection{Damgard-Jurik}
With the Damgard-Jurik method \cite{damgaard2010generalization} we select two large prime numbers ($p$ and $q$) and compute:
\begin{align}
n=pq\\
\phi=(p-1)(q-1)\\
\lambda =\operatorname {lcm} {\phi} 
\end{align}

and where lcm is the least common multiple. We then select a random integer $g$ for:
\begin{align}
\displaystyle g \in \mathbb {Z} _{n^{2}}^{*}
\end{align}

We must make sure that $n$ divides the order of $g$ by checking the following: 
\begin{align}
\mu =(L(g^{\lambda }{\bmod {n}}^{2}))^{-1}{\bmod {n}}
\end{align}

and where $L$ is defined as $L(x)=\frac {x-1}{n}$. The public key is $(n,g)$ and the private key is $\lambda,\mu )$. To encrypt a message ($M$), we select a random $r$ value and compute the ciphertext of:

\begin{align}
c=g^{m}\cdot r^{n} \pmod {n^{s+1}}
\end{align}

and then to decrypt:
\begin{align}
m=L(c^{\lambda }{\bmod n}^{2})\cdot \mu {\pmod n^s} 
\end{align}

If we take two ciphers ($C_1$ and $C_2$), we get:
\begin{align}
C_1 = g^{m_1}\cdot r_1^{n} \pmod {n^{2}} \\
C_2 = g^{m_2}\cdot r_2^{n} \pmod {n^{2}} 
\end{align}

If we now multiply them, we get:

\begin{align}
C_1 \cdot C_2 = g^{m_1}\cdot r_1^{n} \cdot g^{m_2}\cdot r_2^{n} \pmod {n^{2}}\\
C_1 \cdot C_2 = g^{m_1+m_2}\cdot r_1^{n} \cdot r_2^{n} \pmod {n^{2}}
\end{align}

Adding two values requires the multiplication of the ciphers. If we now divide them, we get:

\begin{align}
\frac{C_1}  {C_2} = \frac{ g^{m_1}\cdot r_1^{n}}{g^{m_2}\cdot r_2^{n}} \pmod {n^{2}}   
\frac{C_1}  {C_2} =  g^{m_1-m_2} \frac{r_1^{n}}{r_2^{n}} \pmod {n^{2}} 
\end{align}

Thus, subtraction is equivalent to a division operation. For this, we perform a modular divide operation.

\section{Methodology}
An IPv4 address has four main fields that are defined with integer values in the range 0-255. An example is 12.23.45.67, which is a 32-bit address value and where each of the fields is identified with an 8-bit value. The address then splits into a network part and a host part, such as where 12.23.46.0 might identify a network address, and 0.0.0.67 will identify the host part. Overall, we define the network part with a subnet mask, and where bits that are set to a 1 identify the network part, and where we have a 0, we define the host part. We can then vary the number of 1's from 0 to 32. An example subnet mask where the network part is 24 bits long is 0xffffff00, and which can be represented by 255.255.255.0. This is often identified by the number of bits, such as 1's in the subnet mask, such as '/24'. An example IP address could thus be '192.168.0.10/24', and where the network part is '192.168.0.0' and the host part is '0.0.0.24'.

\subsection{PHE Subtraction method}

Algorithm 1 defines the method used for subtractive homomorphic encryption. In this, we convert the target IP address and the blacklisted network address to 32-bit integer values. The subnet mask then defines the network part to match. Again, this will be an integer value, and where the 1's identify the network part and the 0's will identify the host part. Overall, we are only interested in matching the network part of the target address to the blacklisted network address. We can then create a BFV keypair with a public key ($pk$) and a private key ($sk$). The homomorphic public key is then used to encrypt the target IP address and also the blacklisted address. Once encrypted, we can then perform a homomorphic subtraction. If the network part of the target IP address and the network address match, the result will be an encrypted value of zero. We can then decrypt the result of the homomorphic subtraction, and if we get a zero, we know that the IP address is contained in the blacklist.  

\begin{algorithm}\label{equ1}
  \begin{algorithmic}[1]

\State Set $IP$ with the address to find for an integer
\State Set $Network$ for the blacklisted addresses as an integer
\State Set $Subnet_{mask}$ as the subnet of blacklist
\State Generate $pk,sk$ for homomorphic key pair
\State $IP_e=Enc(IP,pk)$
\State $Blacklist = Network \land Subnet_{mask}$
\State $Blacklist_e=Enc(Blacklist,pk)$
\State $Enc_{diff}=IP_e-Blacklist_e$

\If{$Enc_{diff} = 0$} 
    \State Address is in the blacklist
\Else
        \State Address is not in the blacklist
\EndIf 

\caption{FHE for IP detection} 
\end{algorithmic}
\end{algorithm}

\subsection{Goldwasser–Micali XOR method}
For the Goldwasser–Micali partial homomorphic encryption method, we can use the XOR operation, and where we can XOR the blacklist network address with the network address of the target IP address. This method is defined in Algorithm 2.

\begin{algorithm}\label{equ2}
  \begin{algorithmic}[1]

\State Set $IP$ with the address to find for an integer
\State Set $Network$ for the blacklisted addresses as an integer
\State Set $Subnet_{mask}$ as the subnet of blacklist
\State Generate $pk,sk$ for homomorphic key pair
\State $IP_e=Enc(IP,pk)$
\State $Blacklist = Network \land Subnet_{mask}$
\State $Blacklist_e=Enc(Blacklist,pk)$
\State $Enc_{diff}=IP_e \oplus Blacklist_e$

\If{$Enc_{diff} = 0$} 
    \State Address is in the blacklist
\Else
        \State Address is not in the blacklist
\EndIf 

\caption{FHE for IP detection} 
\end{algorithmic}
\end{algorithm}

\subsection{OpenFHE parameters}
The parameters that need to be set within OpenFHE for BFV are: 

\begin{itemize}
    \item $Scheme$. This defines the scheme to be used. In the case of BFV, this is set to BFVRNS\_SCHEME.
    \item $RDim$. This defines the size of the lattice ring dimension. A typical value for this is 16,384.
    \item $MultDepth$. This is the multiplication depth and is the maximum number of sequential (cascaded) multiplications that are performed on encrypted data before decryption fails due to excessive noise accumulation. 
    \item $PtMod$. This defines the plaintext modulus, and it needs to be a prime number that is larger than the number of bits in the plaintext. 
\end{itemize}

The parameter set needs to support integer values up to 32 bits. A recommended setup for 41-bit resolution for the plaintext ($PtMod$) is set at 35,184,372, 744,193, along with the other parameters defined at \cite{openfhe_test}. The FHE and PHE code is given at \cite{github}. There is a  plaintext constraint that comes from the requirement to support component-wise vector multiplication (instead of polynomial multiplication). In OpenFHE, this is enabled by Chinese Remainder Theorem, and comes down to performing an inverse NTT (Number Theoretic Transform) over the plaintext modulus. The plaintext modulus thus has to support NTT, and where it should be congruent to $1 mod 2 \times N$, where $N$ is the ring dimension \footnote{\url{https://www.microsoft.com/en-us/research/wp-content/uploads/2017/06/sealmanual_v2.2.pdf}} and within more formally in Section 7.4 of \cite{smart2014fully}. The selection of the modulus is provided in the code at \footnote{\url{https://github.com/openfheorg/education/blob/main/examples/modulus_picking/primes.go}}.

\begin{table*}
\begin{center}
\begin{tabular}{ l r r r}
\hline
Method & Key pair (ms) & Encrypt (ms) & Operation and decrypt (ms)\\
\hline
BFV (OpenFHE) \cite{OpenFHE} & 93.1  & 270.2 & 16.4 \\
Paillier \cite{paillier1999public} & 55.7 & 168.6 & 66.3\\
Damgard-Jurik \cite{damgaard2010generalization} & 83.6 & 443.5 & 155.2\\
Okamoto-Uchiyama \cite{okamoto1998new} &  128.9 & 206.9 & 18.0\\
Naccache-Stern \cite{naccache1997new} & 48.2 & 0.2 & 0.6\\
Benaloh & 6.0 & 0.3 & 0.2\\
Goldwasser-Micali \cite{goldwasser2019probabilistic} & 0.3 &  0.5 & 1.6\\
\hline
\end{tabular}
\label{table}\caption{Results for homomorphic operations for IP matching}
\end{center}
\end{table*}

\section{Evaluation}
The coding for fully homomorphic encryption using OpenFHE is defined in the Coding section. The results for a t3.medium instance on AWS are given in Table 1, which includes a comparison with partial homomorphic methods using the PHE Library \cite{PHE}. The time to set up the key pair and the context for the encryption is measured at an average of 93~ms. The greatest overhead is then the time it takes to encrypt the values, which has an average time of around 270~ms. The subtraction and decryption timing then comes in around 16~ms. It can be seen that the encryption process provides the largest overhead in IP address matching. We can see that the Benaloh and Goldwasser-Micali methods are by far the fastest. The BFV method has comparable performance to the PHE methods and is actually faster in the homomorphic encryption operation than Paillier, Damgard-Jurik, and Okamoto-Uchiyama. The Naccache-Stern method also performs well, especially in the encryption and decryption process. Overall, the encryption process tends to have the greatest processing overhead.

The results in Table 2 outline the time to encrypt a number of IP addresses into a data store, and then the total time to search the whole of the data store and find the last matched IP address. As we can see, the encryption time is the most considerable processing overhead.

\begin{table*}
\begin{center}
\begin{tabular}{ l r r r}
\hline
IP addresses & Time to encrypt (secs) & Max time to search (sec)\\
\hline
50 & 6.47 & 1.08\\
100 & 11.55 & 2.48\\
200 & 23.25 & 4.09\\
400 & 125.55 & 8.13\\
800 & 238.93 & 16.57\\
\hline
\end{tabular}
\label{table2}
\caption{Results for fully homomorphic encryption using OpenFHE and matching for a range of random addresses}
\end{center}
\end{table*}

\section{Conclusion}
The increasing requirements for privacy-aware cybersecurity provide opportunities to encrypt data using homomorphic encryption. This paper outlines a method that requires plaintext to support 32 bits and requires a larger plaintext modulus than is used by default in applications. The paper has thus outlined a method that uses the popular OpenFHE library and has fairly reasonable overheads in latency in creating the homomorphic encryption keys and in encrypting, processing, and decrypting data.

\bibliographystyle{IEEEtran}
\bibliography{main}

\section*{Authors}

\begin{wrapfigure}{r}{0.25\textwidth} 
\vspace{-\intextsep}
\centering
\includegraphics[width=0.2\textwidth]{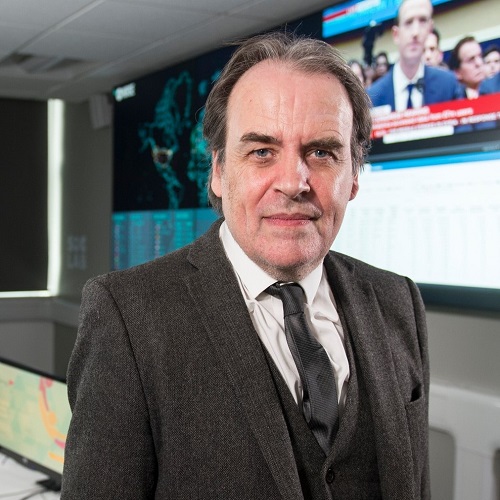}
\end{wrapfigure}

\textbf{William (Bill) J Buchanan OBE FRSE} is a Professor of Applied Cryptography in the School of Computing, Engineering and Built Environment at Edinburgh Napier University. Bill was appointed an Officer of the Order of the British Empire (OBE) in the 2017 Birthday Honours for services to cybersecurity, and,  in 2024, he was appointed as a Fellow of the Royal Society of Edinburgh (FRSE). In 2023, Bill received the "Most Innovative Teacher of the Year" award at the Times Higher Education Awards 2023 (the "Oscars of Higher Education"), and was awarded “Cyber Evangelist of the Year” at the Scottish Cyber Awards in 2016 and 2025. He currently leads the Centre for Cybersecurity, IoT and Cyberphysical, the Blockpass ID Lab, and is the Director of the Scottish Centre of Excellence in Digital Trust and DLT.\\

\begin{wrapfigure}{r}{0.25\textwidth} 
\vspace{-\intextsep}
\centering
\includegraphics[width=0.2\textwidth]{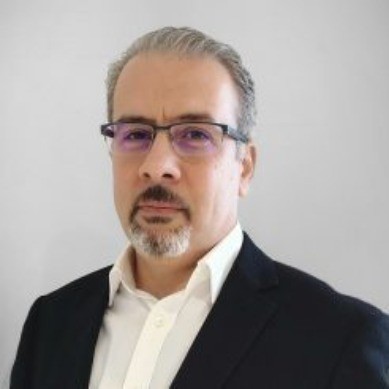}
\end{wrapfigure}

\noindent \textbf{Zakwan Jaroucheh}'s research focuses on decentralisation, with a particular emphasis on blockchain technology, Web3 protocols, crypto assets, and tokenisation. His work has explored blockchain applications in areas such as decentralised authentication, secure digital asset sharing, and combating misinformation, resulting in publications in highly-cited journals and premier international conferences. He has secured significant grants from Innovate UK, The Data Lab, and Scottish Enterprise for his innovative solutions, including a UK patent application for secure crypto asset custody and groundbreaking work in combating impersonation phone call fraud.\\

\begin{wrapfigure}{r}{0.25\textwidth} 
\vspace{-\intextsep}
\centering
\includegraphics[width=0.2\textwidth]{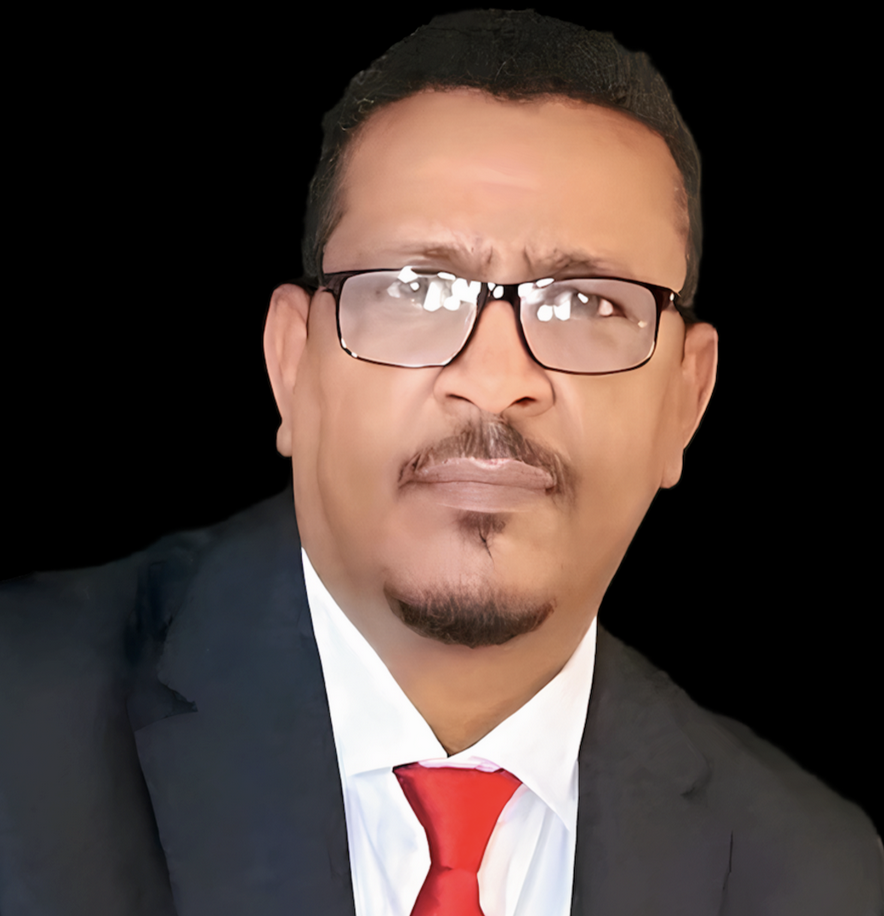}
\end{wrapfigure}

\noindent \textbf{Hisham Ali} graduated with his PhD in 2024, and is now a Research Assistant in the Blockpass ID Lab at Edinburgh Napier University. With a focus on privacy-preserving technologies and ethical hacking, my research is geared towards fortifying digital infrastructures against evolving threats. His involvement with the Hyperledger Community and the Blockpass Identity Lab (BIL) underpins his dedication to distributed ledger technology and its applications in securing data integrity and privacy.  \\

\begin{wrapfigure}{r}{0.25\textwidth} 
\vspace{-\intextsep}
\centering
\includegraphics[width=0.2\textwidth]{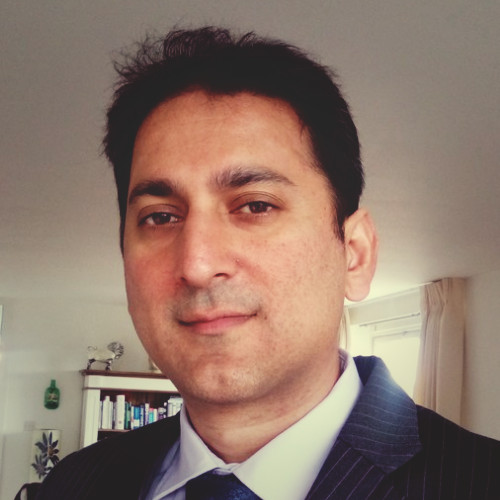}
\end{wrapfigure}

\noindent \textbf{Nanik Ramchandani} is an entrepreneur with over 20 years of experience in the financial services industry, managing wealth, and advising families on global investments.  He is passionate about building solutions and supporting entrepreneurs to solve the biggest challenges of our time. Nanik is a strong believer in \emph{technology for good} and its impact in shaping a better world. Nanik has an MBA from INSEAD and IIM Calcutta. He is a member of Mensa UK and is currently the CEO at LastingAsset - a Trust-tech business on a mission to counter fraud.\\

\begin{wrapfigure}{r}{0.25\textwidth} 
\vspace{-\intextsep}
\centering
\includegraphics[width=0.2\textwidth]{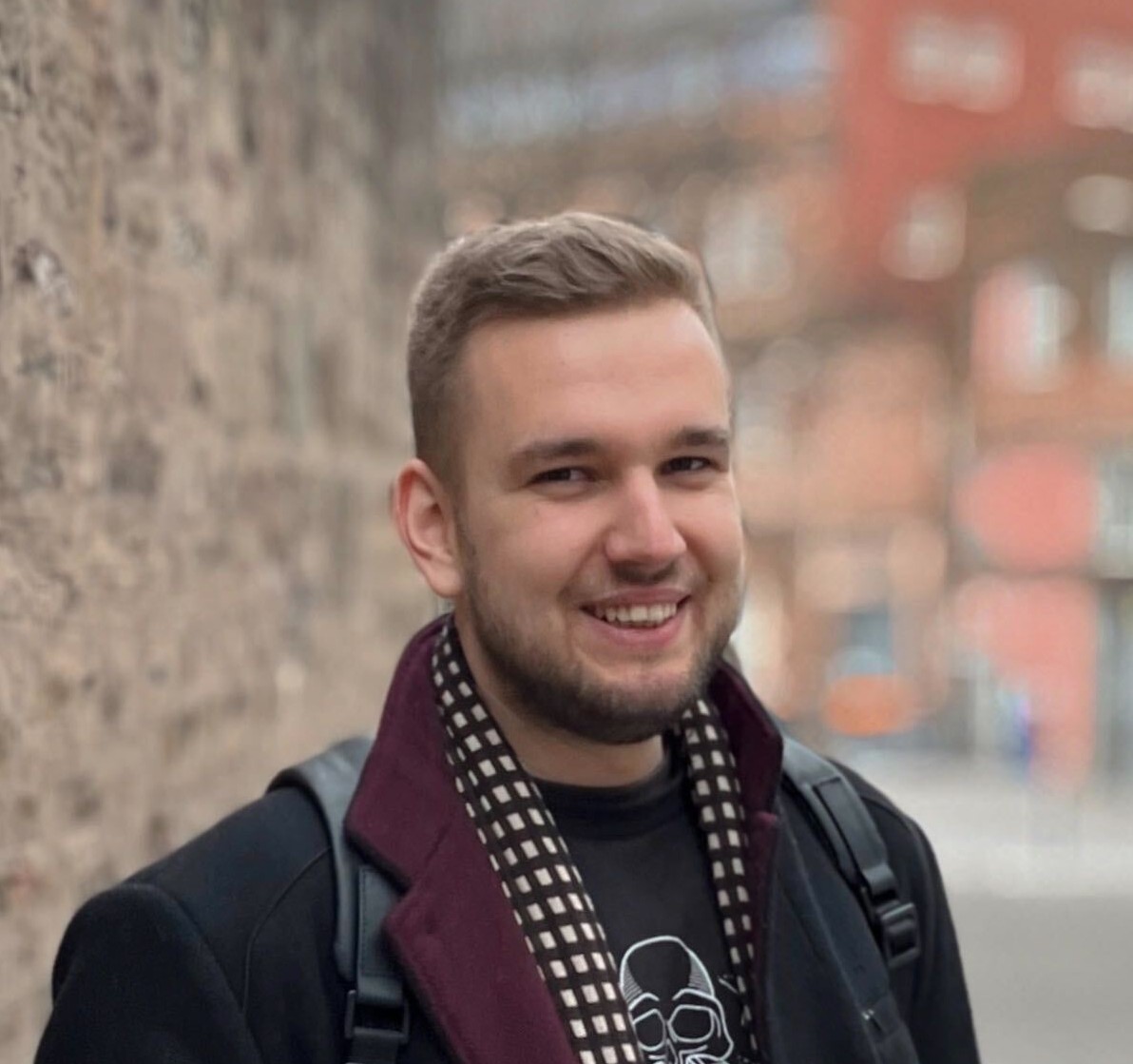}
\end{wrapfigure}

\noindent \textbf{Dmitri Timosenko} is a senior software engineer whose expertise spans full-stack development and defensive cybersecurity, with a particular focus on secure mobile systems. A computing graduate, he has built a career around applying secure-by-design principles across the development lifecycle, from threat modelling to code review. He is best known as the lead developer of the LA Authenticator application, a lightweight multifactor-authentication solution that combines rigorous security guarantees with strong usability. In leading agile teams and mentoring junior engineers, Dmitri champions robust architectures, privacy-preserving design patterns, and continuous integration of security practices. \\

\begin{wrapfigure}{r}{0.25\textwidth} 
\vspace{-\intextsep}
\centering
\includegraphics[width=0.2\textwidth]{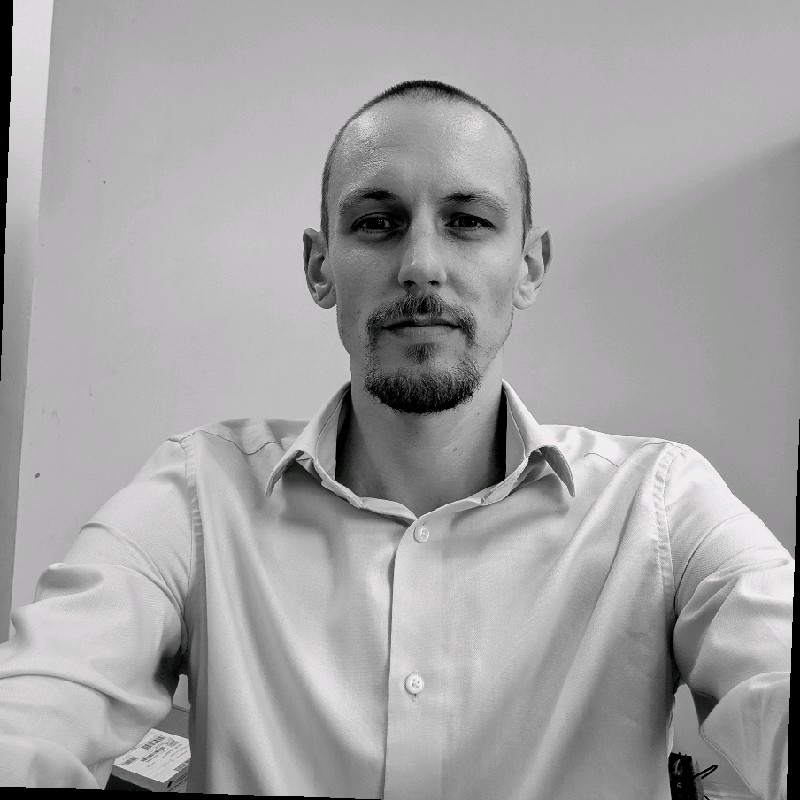}
\end{wrapfigure}

\noindent \textbf{Jamie Gilchrist} is a self-taught cryptography and cybersecurity enthusiast with over 20 years of professional experience in the IT industry. His research interests include applied cryptography, decentralised systems, identity solutions and the intersection of machine learning, privacy and data security. \\

\end{document}